\documentclass[10pt,conference]{IEEEtran}
\IEEEoverridecommandlockouts

\usepackage{siunitx}
\usepackage{cite}
\usepackage{amsmath,amssymb,amsfonts, amsthm}
\usepackage{tikz}
\usetikzlibrary{automata, positioning, arrows,calc}
\usetikzlibrary{quantikz}
\usepackage{algorithmic}
\usepackage{graphicx}
\usepackage{textcomp}
\usepackage{xcolor}
\usepackage{caption} 
\usepackage{subcaption}

\usepackage{hyperref}
\usepackage{mathtools}
\usepackage{multirow}
\usepackage{graphicx}
\usepackage{multicol}
\usepackage{array}
\usepackage{enumitem}

\newtheorem{cor*}{Corollary}

\usepackage{float}

\newcommand{\mpfont}{\scriptsize}

\ifx\noeditingmarks\undefined
    \newcommand{\MPworker}[2]{{\color{#1}\vrule\vrule}{\marginpar{\color{#1}\mpfont #2}}}
\else
    \newcommand{\MPworker}[2]{}
\fi

\usepackage[T1]{fontenc}
\usepackage[english]{babel}
\usepackage{tikz}
\usepackage[utf8]{inputenc}
\usetikzlibrary{shapes.geometric, arrows.meta, positioning}

\begin{document}

\title{Entanglement Meets Reality: \\
A Network Engineering Assessment and Forecast \\ of Rackable Entanglement Sources
}

\makeatletter
\newcommand{\linebreakand}{%
  \end{@IEEEauthorhalign}
  \hfill\mbox{}\par
  \mbox{}\hfill\begin{@IEEEauthorhalign}
}
\makeatother

\author{

\IEEEauthorblockN{Laura d'Avossa$^{*}$, Daniela Salvoni$^{\dagger}$, Angela Sara Cacciapuoti$^{*}$, Marcello Caleffi$^{*}$}

\IEEEauthorblockN{$^{*}$www.QuantumInternet.it research group, University of Naples Federico II, Naples, 80125 Italy.}

\IEEEauthorblockN{$^{\dagger}$Quantech Srl, via Giacinto Gigante 174, 80128 Napoli (NA) Italy}


}

\maketitle

\begin{abstract}
Quantum networks are transitioning from laboratory experiments to real-world deployments, with entanglement as their fundamental resource. Since an entanglement source effectively defines a quantum network, its performance directly impacts the reliability, scalability, and efficiency of future quantum communications. In this work, we investigate rack-mountable plug-and-play entangled-photon sources from a network engineering perspective, shifting the focus from device characterization to deployment-oriented performance evaluation. Building upon an extensive experimental campaign, we assess commercially deployable hardware across multiple operating conditions, evaluate current state-of-the-art capabilities, and provide an outlook on future generations of entanglement sources. We identify and evaluate two key performance indicators (KPIs): multi-photon generation, capturing deviations from ideal single-pair emission, and entanglement quality, quantified through the reconstructed two-qubit density matrix. By combining the measured detected-pair rate with the one-way hashing bound derived from each density matrix, we estimate a lower bound on the achievable distillable-entanglement generation rate, providing a compact metric that captures the trade-off between pair throughput and entanglement quality. Finally, we translate these experimental results into lower bounds on the quantum-memory coherence time required for entanglement distillation, directly linking optical source performance to the hardware requirements of future quantum repeater nodes.

\end{abstract}

\begin{IEEEkeywords}
Quantum Internet; Quantum communications; entanglement; Testbed
\end{IEEEkeywords}

\section{Introduction}
\label{sec:1}

As quantum networks move beyond proof-of-concept demonstrations toward practical deployment, the efficient generation and distribution of high-quality entanglement states becomes a central engineering challenge \cite{RFC9340}.
Indeed, entanglement constitutes the key resource that enables quantum communication protocols and applications beyond the capabilities of classical networks. From an engineering perspective, \textit{an entanglement source defines a quantum network} -- as schematically depicted in Figure~\ref{fig:Figure01} -- and entanglement distribution between nodes of the same network or or between different quantum networks is essential for enabling scalability and advanced quantum communication protocols \cite{TiaWuLi-24, LiuLiWan-24, LiWanMin-22}.
Two main factors can affect the entanglement distribution. One is the quantum channel that can degrade the quality of the distributed ebits.
But the second, and preliminary one, is the entanglement source itself, which determines the quality of the generated quantum states.

\begin{figure}[!t]
    \centering
    \includegraphics[width=\columnwidth]{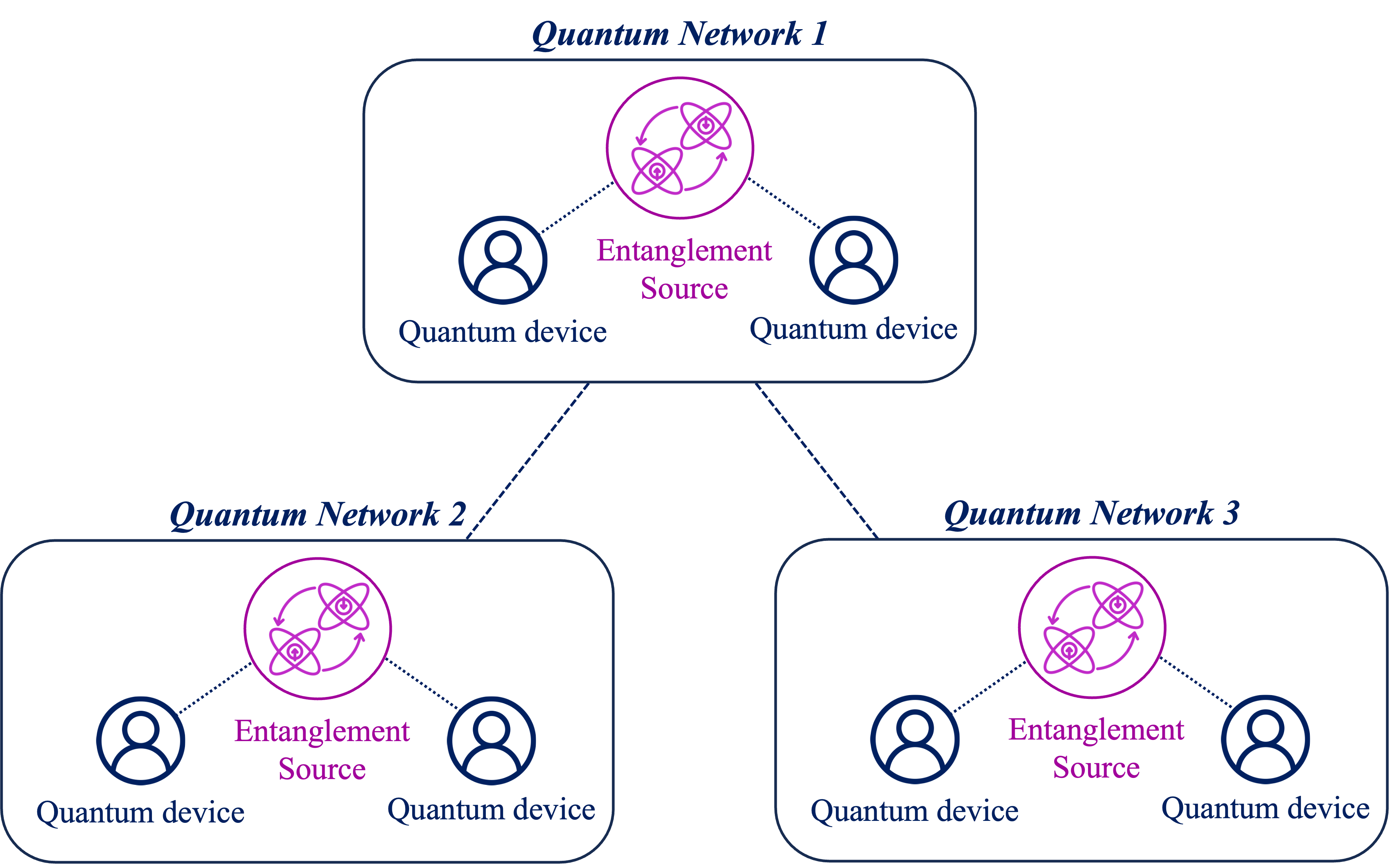}
    \caption{An entanglement source defines a quantum network. The interconnection of multiple quantum networks relies on entanglement distribution.}
    \label{fig:Figure01}
    \hrulefill
\end{figure}

The primary metrics used to characterize an entangled photon source are the pair generation rate and the state fidelity. However, the design of future quantum networks requires going beyond these conventional metrics to understand how they ultimately affect the performance of quantum communication protocols. For instance, achieving high-fidelity entanglement may require long-lived quantum memories to accumulate enough entangled pairs for distillation, whereas a high-brightness source may ultimately deliver a larger amount of usable entanglement.
To address this gap, we introduce a network-oriented characterization framework that maps experimentally measured optical quantities onto deployment-relevant key performance indicators (KPIs). Specifically, we estimate the achievable entanglement throughput and the corresponding minimum quantum-memory coherence time, thereby establishing a direct link between source measurements and the hardware resources required by future quantum networks.

\section{Preliminaries}
\label{sec:02}

\subsection{Entanglement Photon Source}

In this work, we present a comprehensive characterization of a compact rack-mounted Entangled-Photon-pair Source (EPS). 
Indeed, among different technologies proposed as entanglement carriers, optical photons are widely recognized as the most promising candidates and are commonly referred to as \textit{flying qubits} \cite{CacCalTaf-19}. Their weak interaction with the environment enables low-decoherence transmission over existing optical communication infrastructures.
Specifically, in fiber-based quantum networks, quantum communications are typically envisioned in the C-band (across any of the DWDM\footnote{A Dense Wavelength Division Multiplexing (DWDM) a multiplexer/demultiplexer technique that divides the C-band into 40 channels, each with a spectral spacing of 100 GHz.} channels spanning the 1520-1577 nm portion of the optical spectrum), where optical photons in Single-Mode Fibers (SMF) exhibit the lowest transmission loss, approximately 0.2 dB/km \cite{ClaYanWanNejSimJos-24}. 
Within these infrastructures, quantum information can be encoded in different degrees of freedom of single photons, such as polarization\footnote{Other commonly used degrees of freedom for quantum encoding include time-bin \cite{JakPreKau-14, ZhoWanZou-20} and frequency \cite{LuLisGae-23}.}  \cite{YinRenLu-12, TalHesDav-26}, that is one of the most widely employed for quantum communications. Indeed, it enables efficient quantum state preparation, manipulation, and measurement using 
In our measurement campaign, entangled photons are generated using a fiber-coupled source of polarization-entangled photon pairs operating in the $1550 nm$ telecom band\footnote{Producer to be disclosed after blind peer review.}. 
A simplified schematic of the EPS is shown in Figure ~\ref{fig:EPS_setup}.
The source exploits Spontaneous Parametric Down-Conversion (SPDC) \cite{KwiMatei-95}, whereby a pulsed optical pump obtained from an internal continuous-wave (CW) laser generates photon pairs in the maximally entangled Bell state:
\begin{equation}
\ket{\Phi^+}=\frac{1}{\sqrt{2}}\left(\ket{H_sH_i}+\ket{V_sV_i}\right)
\label{eq:state}
\end{equation}
where the subscripts $s$ and $i$ denote the \textit{signal} and \textit{idler}  photons, respectively\footnote{The EPS supports multiple signal-idler wavelength-channel pairs through an internal filtering system. In this work, the characterization is performed using the 24–44 and 23–45 signal-idler channel pairs, thus eliminating the need for external DWDM components.}.
The EPS provides a number of configurable operating parameters that can affect the quality of the generated entanglement. 
To this end, two fundamental operating parameters of the source are systematically varied to investigate their impact on the quality of the generated entanglement.
The first, referred to as \textit{frequency}, controls the repetition rate of the optical pulses generated by modulating the internal continuous pump laser. The second, referred to as \textit{VOA} (Variable Optical Attenuator), controls the attenuation applied to these pump pulses. By tuning these two parameters, the photon-pair source can be characterized under different operating conditions.

\begin{figure}[!t]
    \centering
    \includegraphics[width=\columnwidth]{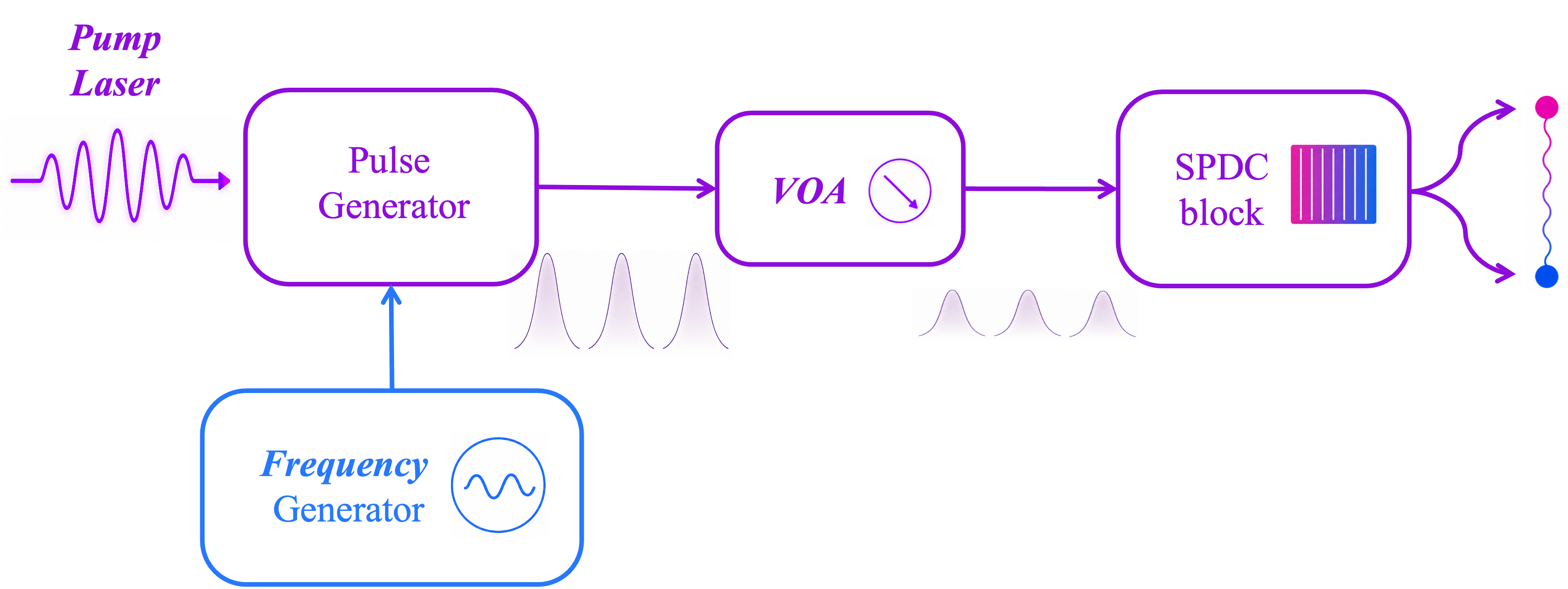}
    \caption{Simplified schematic of the operating principle of the EPS. A CW Pump Laser is converted into a train of optical pulses by the pulse generator, whose repetition rate is set by the \textit{frequency} parameter. The pulse energy is subsequently adjusted through a \textit{VOA} (Variable Optical Attenuator) before pumping the nonlinear crystal, where polarization-entangled photon pairs are generated via SPDC.}
    \label{fig:EPS_setup}
    \hrulefill
\end{figure}

\subsection{Fock States}

The quantum state of light can be represented in different bases depending on the physical quantity of interest. For the characterization of photon-pair sources, the most convenient representation is the Fock basis, whose basis vectors describe states with a well-defined number of photons \cite{Loudon2000, WallsMilburn2008}. A Fock state is denoted by $\ket{n}$, where the integer $n\in\mathbb{N}$ specifies the exact number of photons in the quantum state.

Since the entanglement generation process of our EPS is a probabilistic process, a single pump pulse does not generate deterministically one entangled photon-pair. Instead, the emitted quantum state is described by a superposition of different photon-number components, including the vacuum state (corresponding to the absence of photon-pair generation), the single-pair state, and higher-order multi-pair contributions. This quantum state can be expressed in the Fock basis as \cite{MeySilMig-20, Lvovsky2015}:
\begin{align}
\ket{\psi}
&=
\sqrt{P_0}\ket{0,0}
+\sqrt{p_g}\ket{1,1}
+\sqrt{P_2}\ket{2,2}
+\cdots \nonumber\\
&=
\sum_{n=0}^{\infty}\sqrt{P_n}\ket{n,n}.
\label{eq:multiphoton}
\end{align}
where $\ket{n,n}\equiv\ket{n}_{s}\ket{n}_{i}$ denotes the state containing $n$ photons in the signal ($s$) and $n$ photons in the idler ($i$), while $P_n$ is the probability of generating the corresponding $n$-photon-pair component.
For quantum communication, the desired operating condition is the one in which the source emits exactly one photon pair per laser pulse, corresponding to the $\ket{1,1}$ Fock component of Eq.~\ref{eq:multiphoton}. In other words, the probability of generating a single photon-pair ($p_g$) should dominate over both the vacuum contribution ($P_0$) and the higher-order terms ($P_n$) with ($n>1$).
Indeed, this single-pair contribution encodes polarization entanglement, while higher-order Fock components constitute an undesired source of noise that limits the performance of photon-pair sources\footnote{The vacuum states are not detected, so they do not affect the entanglement fidelity}.
Figure~\ref{fig:fock_pulses} schematically illustrates the photon-number generated associated with successive pump pulses\footnote{Let us notice that current entangled photon-pair sources operate in a regime where the probability of generating a useful photon pair per pump pulse is extremely small.}.
As will become evident in the following sections, increasing the classical pump power that drives the probabilistic entanglement generation process increases the probability of successfully generating the target $\ket{11}$ state. At the same time, however, it also raises the likelihood of unwanted multiphoton-pair generation. Using our Fock-state formalism, we evaluate not only the average number of generated photons but also characterize the intrinsic trade-off between a higher success probability for the target state and the unwanted increase in multiphoton contributions.

\begin{figure}[!t]
    \centering
    \includegraphics[width=\columnwidth]{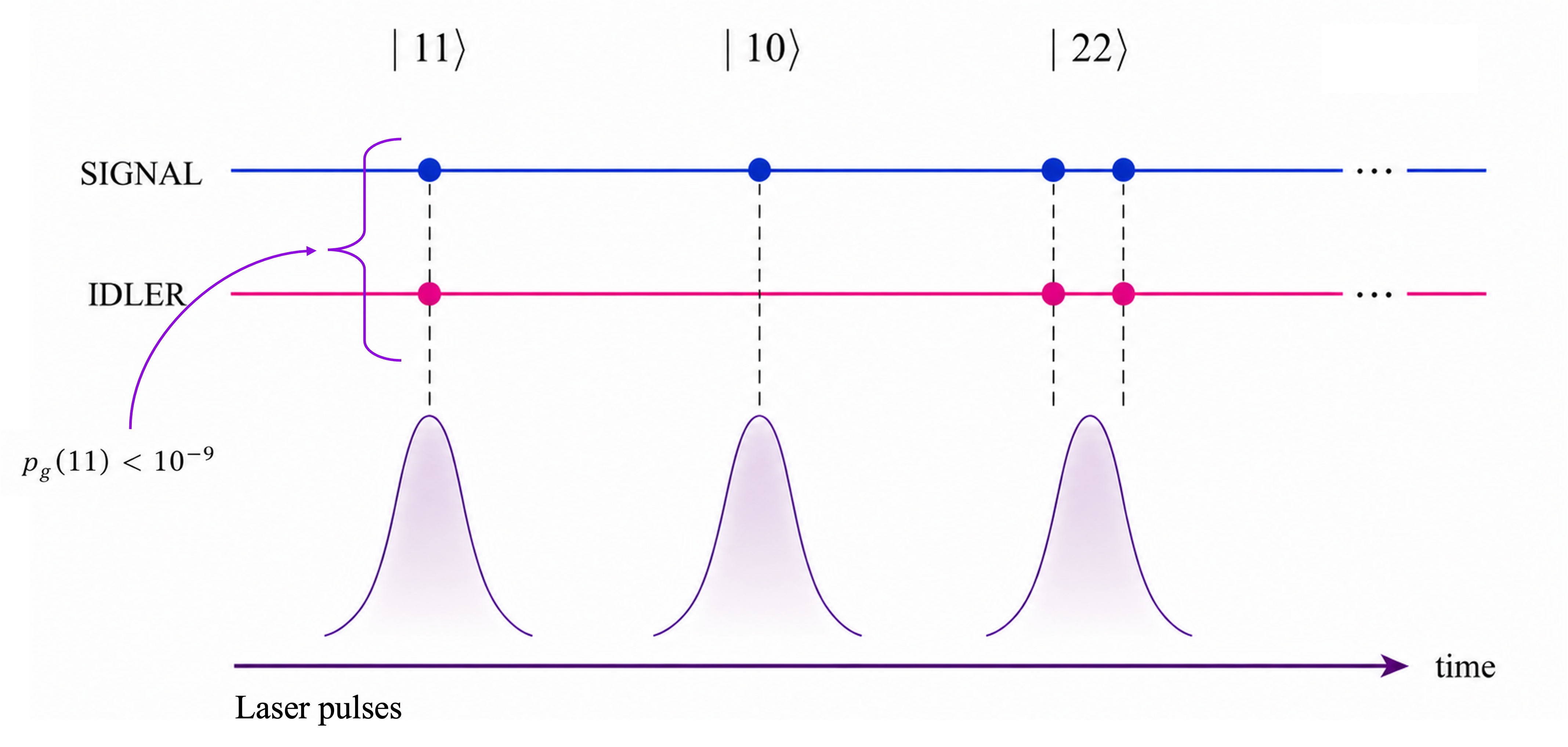}
    \caption{Schematic representation of the possible Fock-state outcomes associated with successive pump pulses in the EPS. The desired event corresponds to the single-pair state $\ket{1,1}$, generated with probability $p_g$, while higher-order components such as $\ket{2,2}$ represent multi-pair emission events.}
    \label{fig:fock_pulses}
    \hrulefill
\end{figure}

\subsection{Entanglement Fidelity and Entanglement Distillation}

In realistic quantum communications scenarios, imperfections in state preparation, transmission noise, and decoherence progressively degrade the entangled state. The quality of an entangled state is commonly quantified through its \textit{entanglement fidelity}, defined as the overlap between the generated state $\rho$ and a target maximally entangled Bell state $\ket{\Phi}$ as: $F(\rho)=\bra{\Phi}\rho\ket{\Phi}$. A fidelity close to unity indicates that the state closely approximates the desired maximally entangled state, whereas lower values correspond to progressively degraded quantum correlations.
To counteract the effects of noise, \textit{entanglement distillation} employs local operations and classical communication (LOCC) to probabilistically extract a smaller number of high-fidelity entangled pairs from a larger ensemble of imperfect ones \cite{nielsen2010quantum, Bennett-96}. The ultimate performance achievable by entanglement distillation protocols is quantified by the distillable entanglement, $E_D(\rho)$, defined as the maximum asymptotic rate at which Bell pairs can be extracted from many copies of a bipartite state $\rho$ using LOCC \cite{HorHorHor-98}. Consequently, the distillable entanglement quantifies the amount of entanglement that can be asymptotically converted into maximally entangled pairs by LOCC, thereby providing an operational measure of the useful entanglement contained in a quantum state.

\section{Multiphoton Analysis}

This section presents the experimental characterization of the EPS multiphoton emission under different operating conditions. Specifically, we investigate its dependence on the frequency and VOA settings, compare different EPS units and DWDM channel pairs, and assess its temporal stability.
Since multiphoton emission is one of the main non-idealities affecting entangled photon sources, its characterization provides a fundamental figure of merit for assessing the source quality and serves as the basis for the subsequent distillable entanglement analysis.

\subsection{Experimental setup}

\begin{figure}[t!]
    \centering
    \includegraphics[width=\columnwidth]{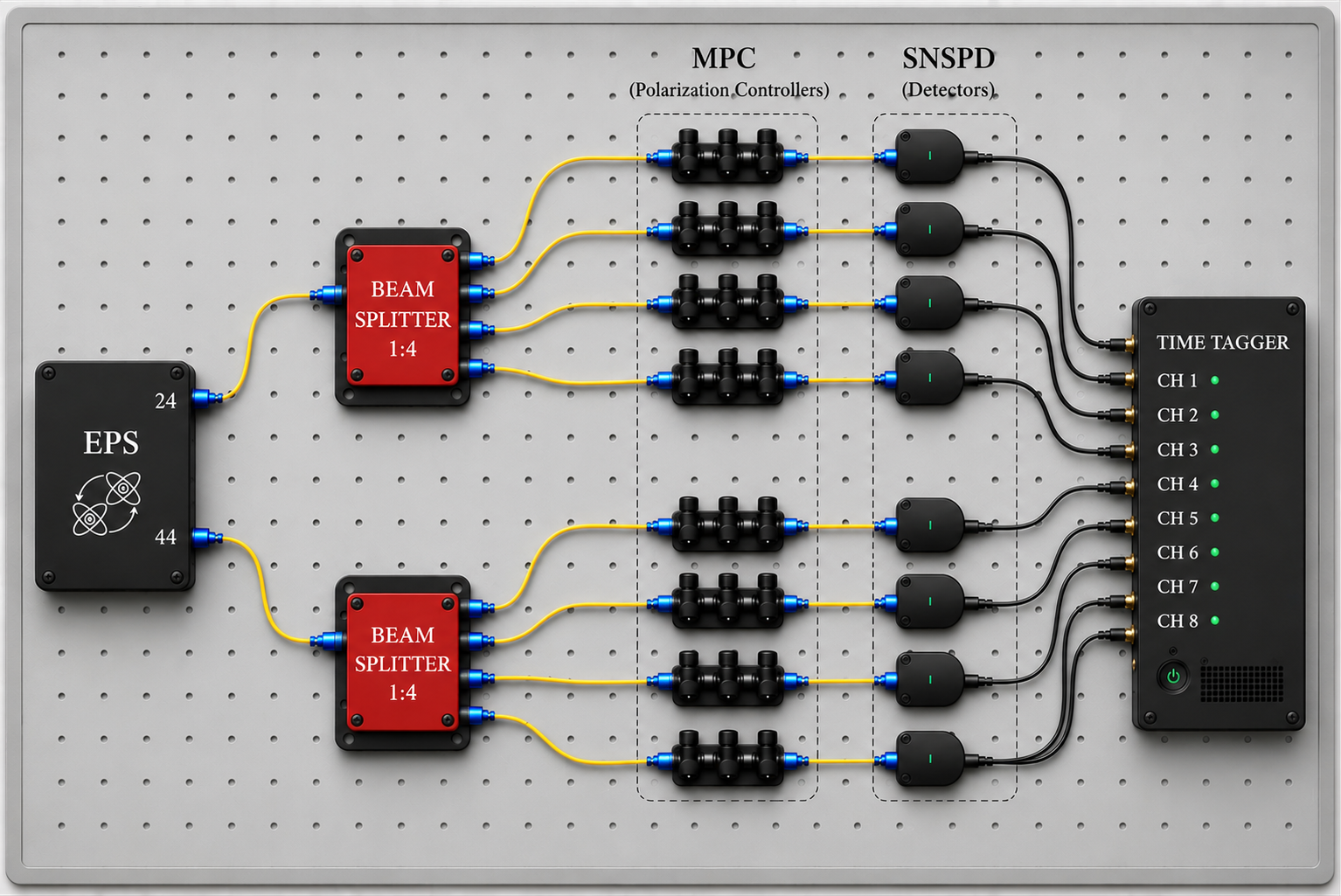}
    \caption{The entangled photon source (EPS) generates signal and idler photons at ITU channels 24 and 44, respectively. Each output channel is connected to a $1\times4$ fiber beam splitter, whose four outputs pass through manual polarization controllers (MPCs) before being coupled to independent superconducting nanowire single-photon detectors (SNSPDs). The eight detectors implement a spatially multiplexed pseudo photon-number-resolving (PNR) detection scheme, while the resulting detection events are acquired by a time tagger to reconstruct the coincidence patterns used for photon-number estimation.}
    \label{fig:setup}
    \hrulefill
\end{figure}

\begin{figure*}[t!]
    \centering
    \includegraphics[width=0.9\textwidth]
    {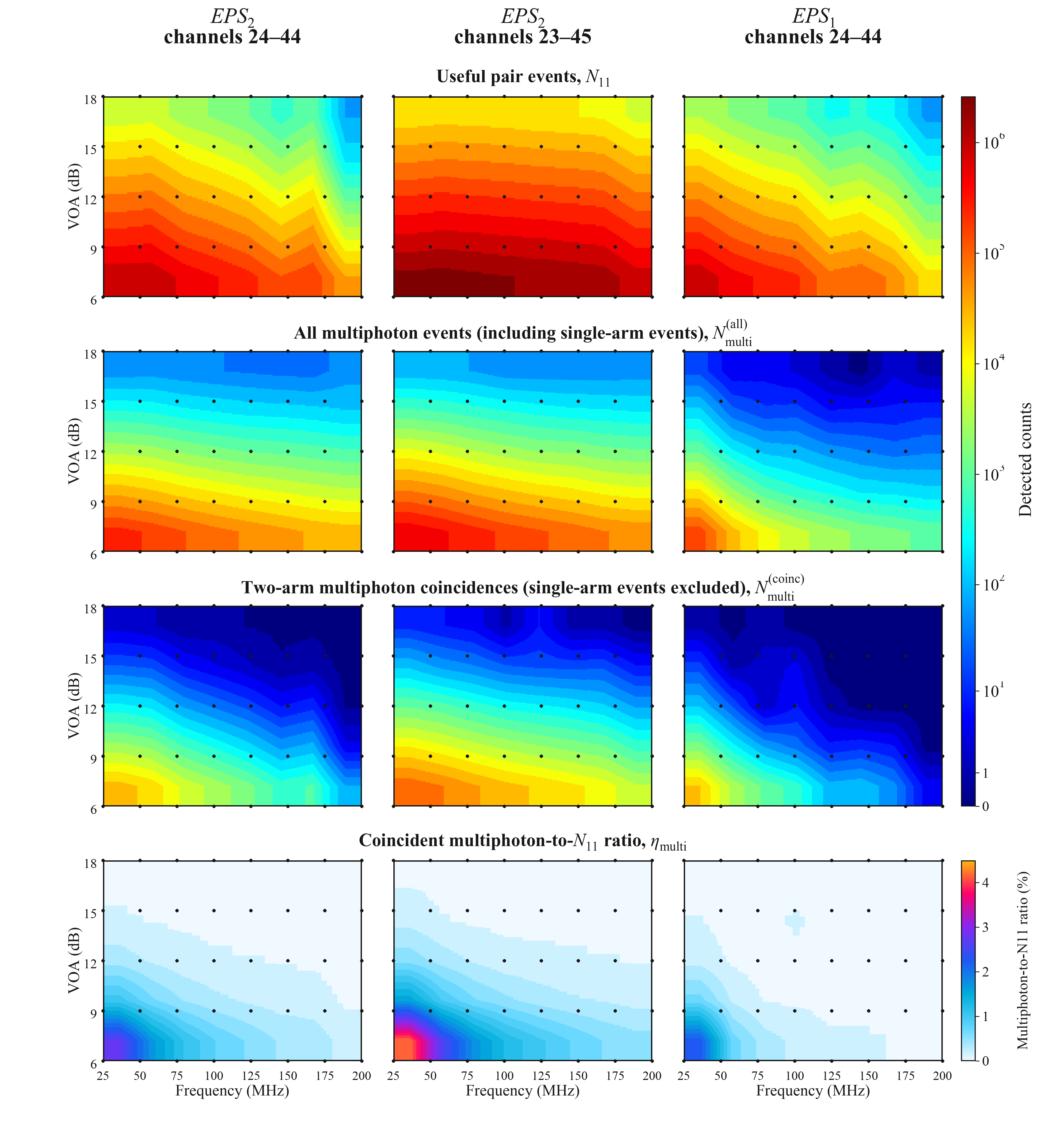}
    \caption{Operating maps of the useful $N_{11}$ events, all detected
    multiphoton events $N_{\mathrm{multi}}^{(\mathrm{all})}$, two-arm multiphoton coincidences obtained by excluding
    single-arm events $N_{\mathrm{multi}}^{(\mathrm{coinc})}$, and the corresponding multiphoton-to-$N_{11}$ ratio $\eta_{\mathrm{multi}}$.
    Columns compare the investigated source and ITU channels. The
    first three rows share a logarithmic count scale, whereas the fourth row
    uses an independent percentage scale.}
    \label{fig:multiphoton-definitions}
    \hrulefill
\end{figure*} 
The experimental characterization relies on a spatially multiplexed detection architecture, which enables the reconstruction of the photon-number statistics at the outputs of the EPS.
The goal is to infer the photon number at each output of the EPS (signal and idler) by implementing a spatially multiplexed detection scheme. To this end, each output is connected to a $1:4$ beam splitter\footnote{The beam splitter output ports exhibit nearly uniform splitting ratios, typically between $24\%$ and $26\%$ per port.}. 
Each beam splitter output is connected to a Superconducting Nanowire Single-photon Detector (SNSPD). Before each detector, a Manual Polarization Controller (MPC) is used to optimize the polarization state of the incoming photons, thereby maximizing the detection efficiency of the corresponding detector. Overall, the setup comprises eight SNSPDs \cite{NatTanHad-12}, fabricated and installed in a rack-mounted configuration\footnote{The cryogenic stage operates at a temperature of $2.2\,\mathrm{K}$ and a pressure of approximately $10^{-6}\,\mathrm{mbar}$. Under these conditions, together with the detector characterization performed by tuning the bias current, at $1550\,\mathrm{nm}$ the system achieves a system detection efficiency (SDE) of about $50\%$, a timing jitter of $50\,\mathrm{ps}$, and a dark count rate (DCR) of approximately $100\,\mathrm{cps}$. Producer to be disclosed after blind peer review.}. 
Four SNSPDs are connected to each beam-splitter, such that each EPS output port is monitored by four independent detection channels. 
Multiple photons arriving at the same beam-splitter are probabilistically distributed among the outputs, allowing the number of coincident detection events to be used as an estimate of the incident photon number. This realizes a pseudo photon-number-resolving (PNR) detection scheme based on spatial multiplexing.
The measurements between the detectors are analyzed using a 
time tagger with resolution of 8 ps.
The complete system setup is shown in Figure~\ref{fig:setup}.
The characterization is initially performed on the ITU 24–44 DWDM channel pair. Subsequently, the same measurement procedure is repeated using the ITU 23–45 channel pair. Finally, the entire characterization is carried out on a second rack-mountable EPS source of identical design. This allows a direct comparison of the multi-photon performance across different wavelength-channel pairs and different sources, providing an assessment of the reproducibility and stability of the implemented system\footnote{In the remainder of this work, we refer to the EPS source used to investigate different DWDM output-channel pairs as $EPS_2$, while $EPS_1$ denotes the source employed to complete the validation of the proposed characterization procedure.}.

\subsection{Experimental results}

We define the total number of multiphoton events as\footnote{The summation extends over all experimentally resolved photon-number outcomes, corresponding to up to four detected photons per output arm.}:
\begin{equation}
N_{\mathrm{multi}}^{(\mathrm{all})}
=
\sum_{i\geq2\,\lor\,j\geq2}N_{ij}.
\end{equation}

This quantity includes every event in which at least one output arm contains more than one detected photon. It therefore accounts for both coincidence events (e.g., $12$, $21$, $22$, $31$, $32$ etc.) and single-arm events (e.g., $20$, $30$, $02$, $03$ etc.).  It is important to notice that only events producing detections on both output arms contribute to the coincidence dataset considered in the subsequent polarization analysis. We therefore define the number of coincident multiphoton events as:
\begin{equation}
N_{\mathrm{multi}}^{(\mathrm{coinc})}
=
\sum_{\substack{i\geq1,\;j\geq1\\i\geq2\,\lor\,j\geq2}}
N_{ij}.
\end{equation}
This quantity retains only multiphoton coincidence events that impact on the reconstructed quantum state and the corresponding performance metrics (Sec.~\ref{sec:Distillable Entanglement Analysis}).
\begin{figure}[t!]
    \centering
    \includegraphics[width=\columnwidth]{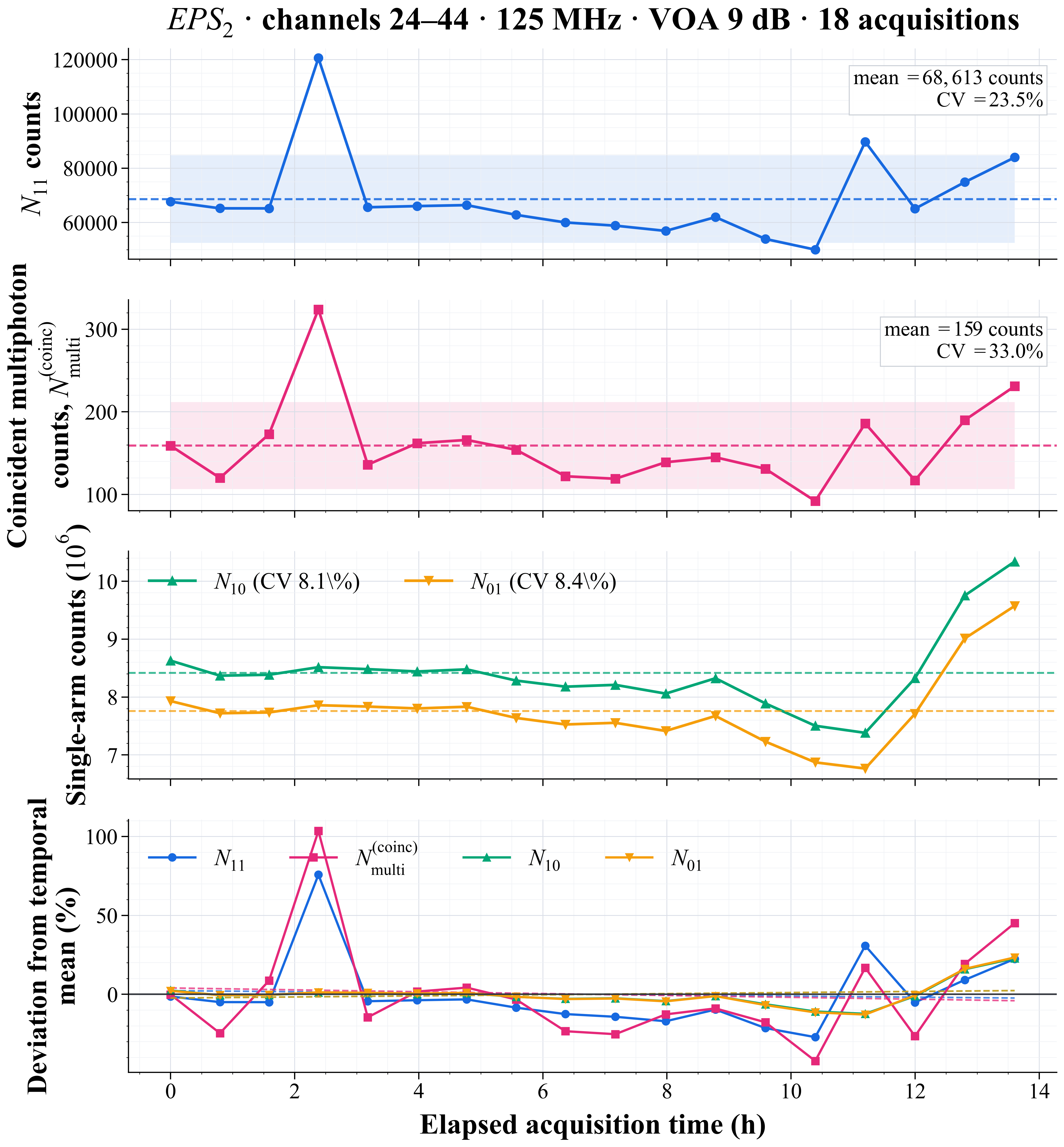}
    \caption{Temporal evolution of the detected emission from $EPS_2$,
    channels 24-44, at $125$~MHz and VOA $9$~dB. Dashed horizontal
    lines indicate temporal means, shaded regions represent one standard
    deviation, and the final panel reports deviations from the
    corresponding mean values.}
    \label{fig:multiphoton-drift}
    \hrulefill
\end{figure}
Figure~\ref{fig:multiphoton-definitions} compares $N_{11}$,
$N_{\mathrm{multi}}^{(\mathrm{all})}$, and
$N_{\mathrm{multi}}^{(\mathrm{coinc})}$ for all the investigated EPSs and DWDM channel pairs as a function of frequency and VOA.
All three quantities generally increase as the VOA is reduced. Their dependence on the frequency is less uniform and exhibits local non-monotonic variations across the investigated EPSs and DWDM channel pairs. Specifically, $EPS_2$ with channels 23-45 exhibits the largest absolute $N_{11}$ and $N_{\mathrm{multi}}^{(\mathrm{coinc})}$ across most of the investigated operating points, defined by the different combinations of VOA and frequency. Conversely, $EPS_1$ generally exhibits lower
$N_{\mathrm{multi}}^{(\mathrm{coinc})}$ values. At high VOA, no coincident multiphoton events are detected at several operating points within the finite acquisition interval.

To quantify how rapidly
$N_{\mathrm{multi}}^{(\mathrm{coinc})}$ grows with respect to the useful $N_{11}$ events, we analyse the quantity:
\begin{equation}
\eta_{\mathrm{multi}}
=
100\,
\frac{N_{\mathrm{multi}}^{(\mathrm{coinc})}}{N_{11}}.
\end{equation}
Basically, an increase in $\eta_{\mathrm{multi}}$ indicates that two-arm multiphoton coincidences grow faster than $N_{11}$. Figure~\ref{fig:multiphoton-definitions} shows that, although reducing the VOA increases the overall number of generated photon pairs, it also leads to a progressively larger fraction of coincidence events originating from higher-order emissions. Thus, $\eta_{\mathrm{multi}}$ provides a direct measure of how efficiently an increase in brightness translates into useful $N_{11}$ events rather than additional multiphoton coincidences.
The maximum ratios are approximately $4.24\%$ for $EPS_2$ channels 23-45, $2.76\%$ for $EPS_2$ channels 24-44, and $2.28\%$ for $EPS_1$ channels 24-44. These results indicate that the choice of the source operating point requires balancing source brightness against the increasing relative contribution of higher-order emissions.

It is important to emphasize that $\eta_{\mathrm{multi}}$ is defined from detected coincidence events and therefore depends not only on the intrinsic multi-photon emission of the EPS but also on the characteristics of the detection system, in particular its efficiency. For instance, events classified as single-arm multiphoton detections (e.g., $N_{20}$ or $N_{02}$) may actually originate from two-arm multiphoton emissions in which photons in one output arm remain undetected because of the detectors efficiency.

\subsection{Temporal Stability of Multiphoton Emission}

The temporal stability of the detected emission is evaluated for $EPS_2$, channels 24-44, at a fixed operating point of $125$~MHz and $9$~dB.
Figure~\ref{fig:multiphoton-drift} reports $N_{11}$, the coincident
multiphoton counts $N_{\mathrm{multi}}^{(\mathrm{coinc})}$, the single-arm counts $N_{10}$
and $N_{01}$, and their deviations from the corresponding temporal means over 18 repeated acquisitions.

The results show that the multiphoton contribution exhibits larger relative fluctuations than the desired $N_{11}$ events. By comparison, the single-arm counts are more stable, with coefficients of variation close to $8\%$.

The common peaks visible in $N_{11}$ and $N_{\mathrm{multi}}^{(\mathrm{coinc})}$ suggest that part of their variability originates from fluctuations in the overall emission. Nevertheless, the linear trends are
weak: 
the measurements do not provide evidence of a systematic monotonic drift over the observed interval. The recorded variability is instead dominated by short-term fluctuations and isolated peaks.

\section{Distillable Entanglement Analysis}
\label{sec:Distillable Entanglement Analysis}

A quantum state encoded in polarization can be reconstructed using Quantum State Tomography (QST), a procedure that  reconstructs the density matrix of the state from measurements performed in different polarization bases \cite{AltJefKwi-05}.
In our experiment, QST is performed on entangled photon pairs generated by the EPS. The reconstructed quantum state is obtained from coincidence measurements, i.e., detection events in which at least one photon is registered on each output arm ($i\geq1,\;j\geq1$).
The reconstructed density matrix can then be used to evaluate the fidelity with the target Bell state Eq.\ref{eq:state}, the distillable entanglement, and other quantities characterizing the quality of the generated entangled state.

\subsection{Experimental setup}

The experimental setup adopted for the QST is shown in Figure~\ref{fig:setup2}. The signal and idler photons generated by the EPS at ITU channels 24 and 44 are analyzed using two Polarization Analyzers (PAs), one for each output. Each PA is followed by an MPC, which compensates for polarization changes introduced by the optical fibers, before the photons are detected by two SNSPDs, one for each channel. Finally, the electrical outputs of the SNSPDs are connected to a time tagger, which records the photon arrival times and identifies coincidence events between the signal and idler detection channels. 

\begin{figure}[t!]
    \centering
    \includegraphics[width=\columnwidth]{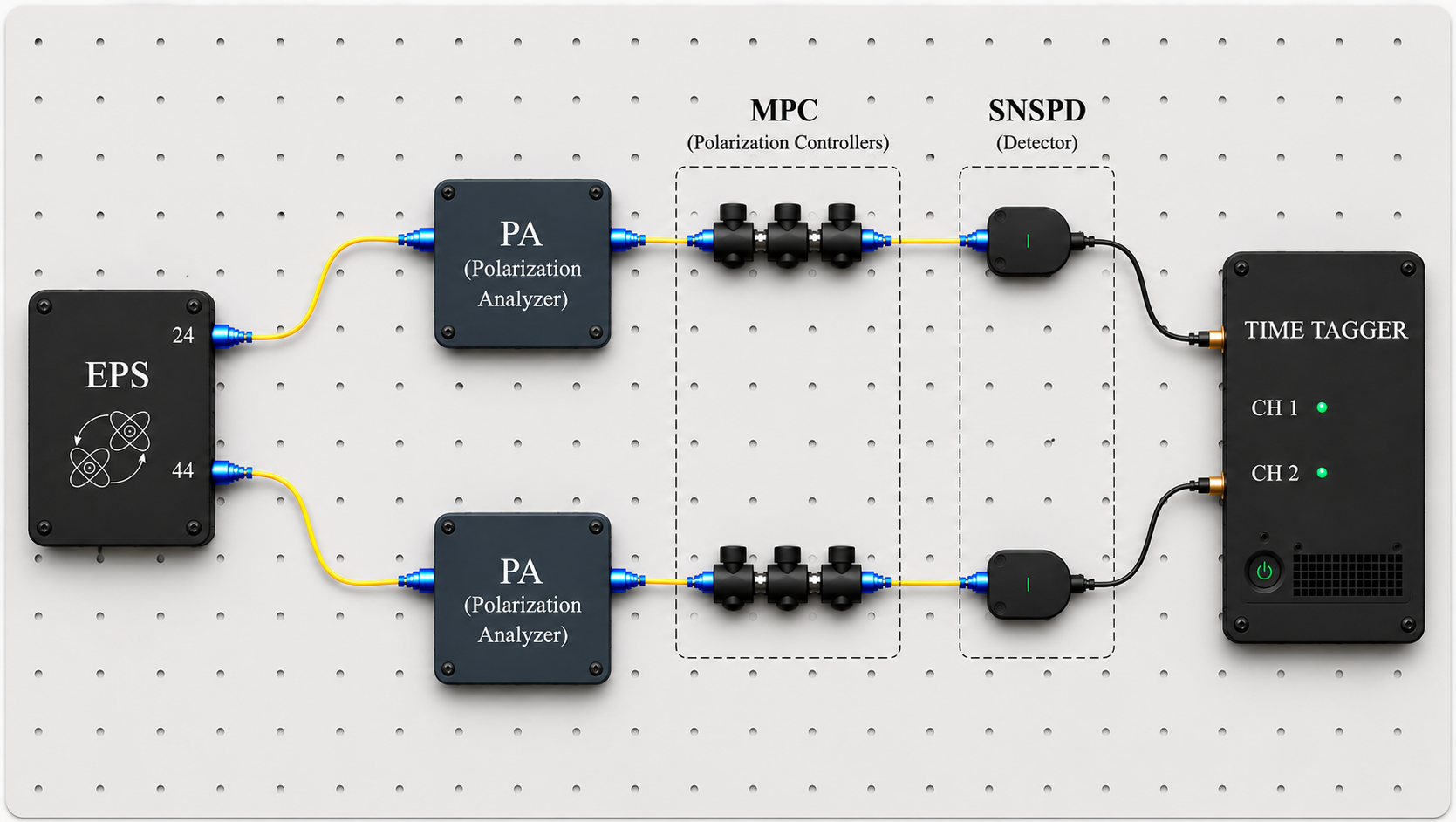}
    \caption{Experimental setup for the fidelity characterization of the entangled photon source. The EPS generates polarization-entangled photon pairs at ITU channels 24 (signal) and 44 (idler). Each output is analyzed by an independent polarization analyzer (PA), followed by a manual polarization controller (MPC) to optimize the polarization at the detector input. The photons are detected by two superconducting nanowire single-photon detectors (SNSPDs), whose electrical outputs are connected to a time tagger for coincidence acquisition.}
    \label{fig:setup2}
    \hrulefill
\end{figure}

\subsection{Experimental results}
\label{sec:Experimental reuslts}

\begin{figure*}[t!]
    \centering

    \begin{subfigure}[t]{\textwidth}
        \centering
        \includegraphics[width=0.9\textwidth]{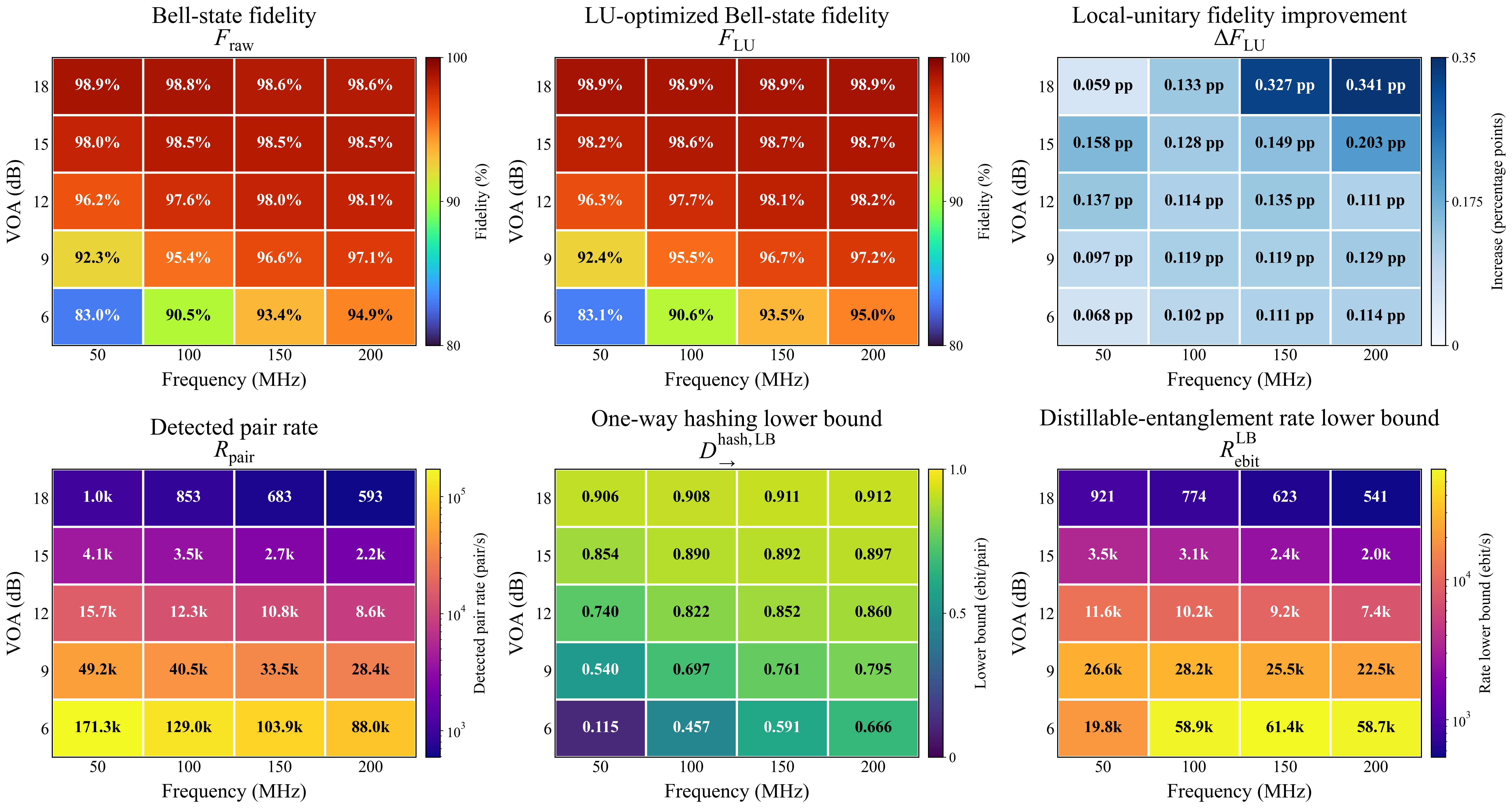}
        \caption{$EPS_2$}
        \label{fig:fidelity_eps2}
    \end{subfigure}

    \vspace{0.5cm}

    \begin{subfigure}[t]{\textwidth}
        \centering
        \includegraphics[width=0.9\textwidth]{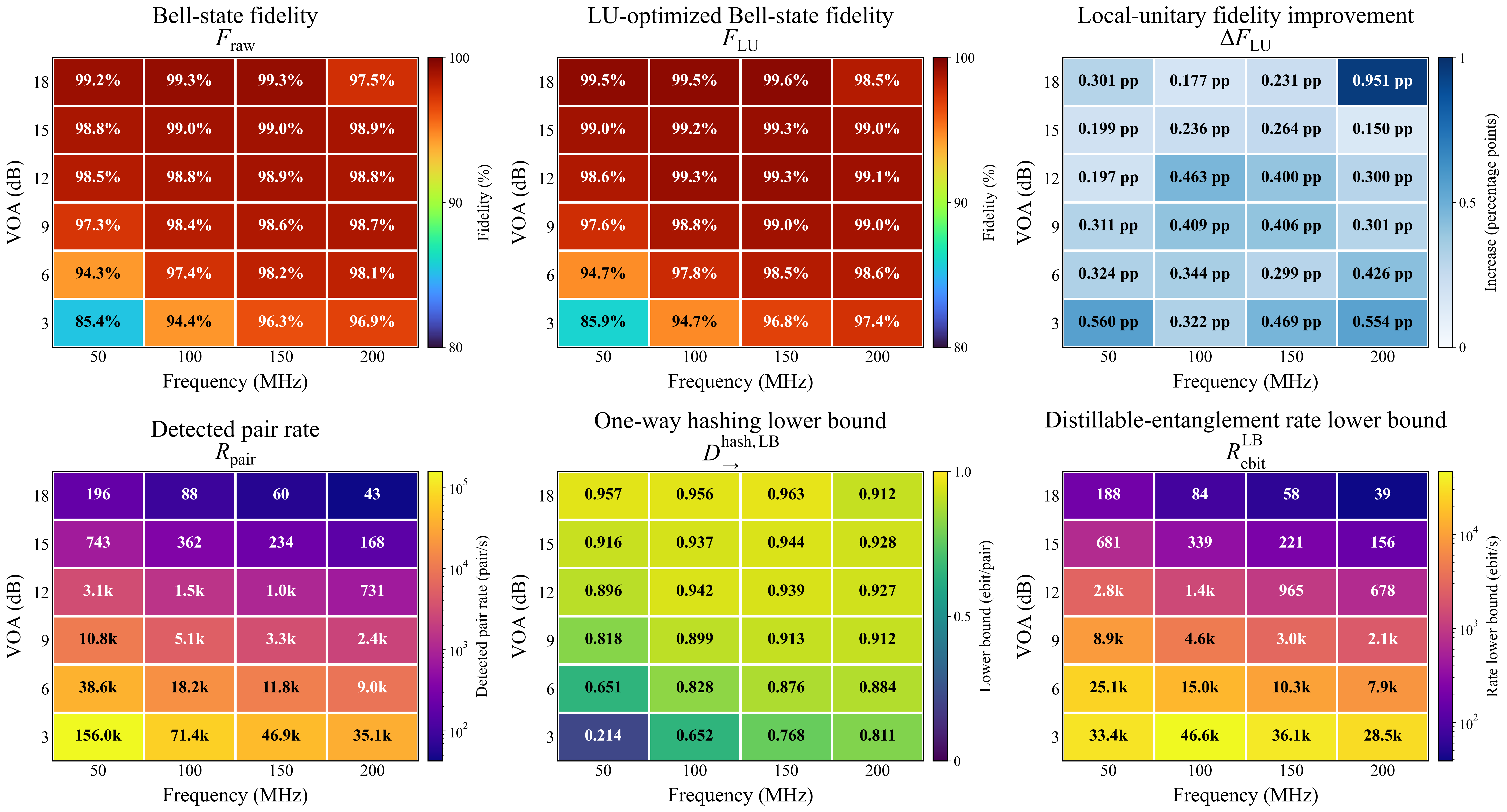}
        \caption{$EPS_1$}
        \label{fig:fidelity_eps1}
    \end{subfigure}

    \caption{Experimental characterization of $EPS_2$ (top) and $EPS_1$ (bottom), channels 24--44, as a function of repetition frequency and VOA. For each source, the upper row shows the Bell-state fidelity, the locally optimized fidelity, and their difference. The lower row shows the detected pair rate, the one-way hashing lower bound, and the resulting distillable-entanglement rate lower bound.}
    \label{fig:fidelity}
    \hrulefill
\end{figure*}

\begin{figure}[t!]
    \centering
    \includegraphics[width=\linewidth]{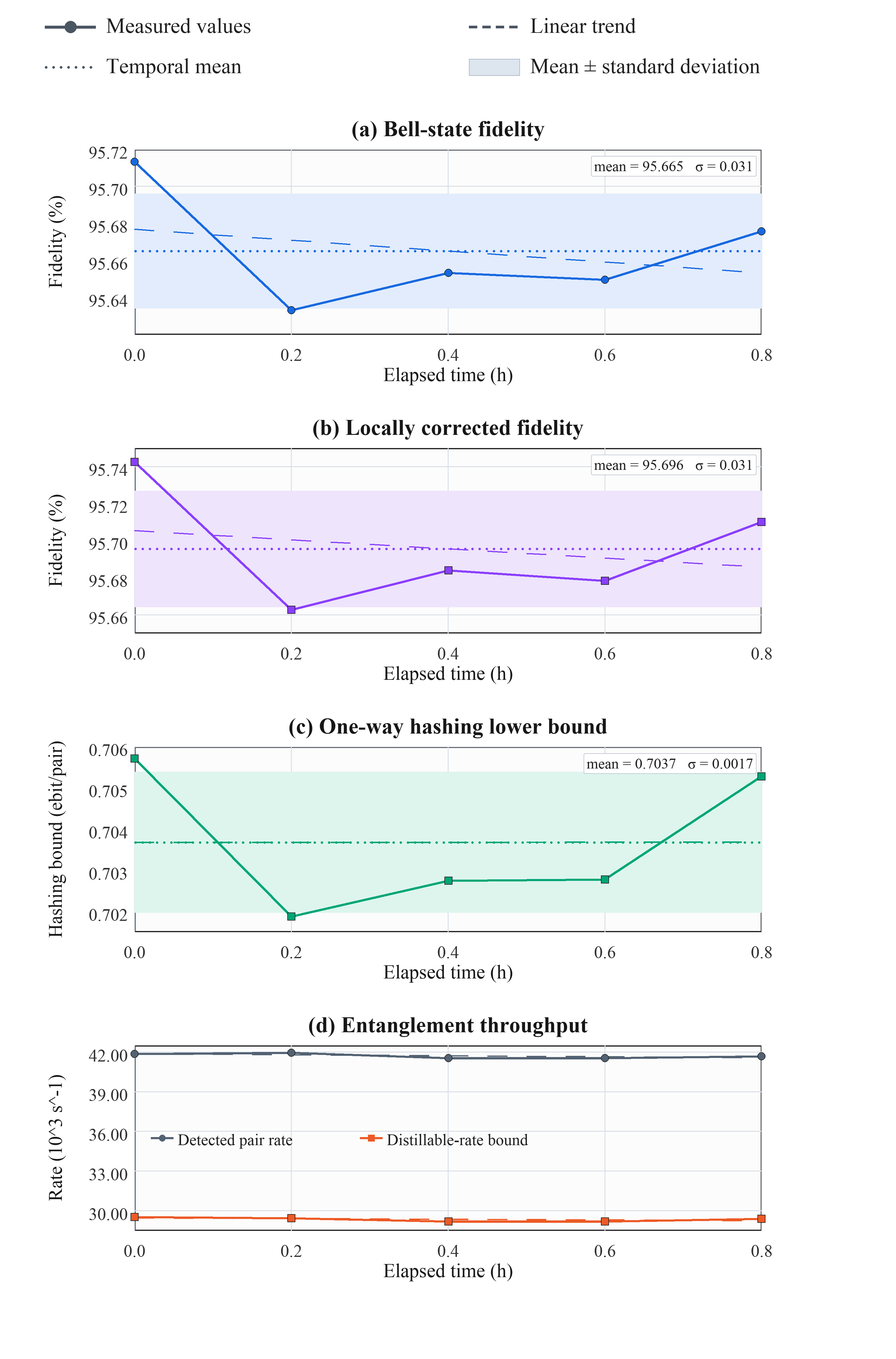}
    \caption{Stability of the entanglement metrics measured for
    $EPS_2$, channels 24--44, at $100$~MHz and VOA $9$~dB. Solid lines with
    markers show the measured values, dashed lines show linear fits,
    dotted horizontal lines indicate temporal means, and shaded regions
    represent one standard deviation around the mean.}
    \label{fig:fidelity-temporal-stability}
    \hrulefill
\end{figure}

At each operating point, namely for each combination of VOA and frequency, the generated state is reconstructed through complete $6\times6$ polarization tomography.

Figure~\ref{fig:fidelity} summarizes the resulting performance metrics for the two entangled-photon-pair sources under same channel configuration (ITU channels 24–44). Figure~\ref{fig:fidelity_eps2} reports the results for $EPS_2$, whereas Figure~\ref{fig:fidelity_eps1} shows the corresponding measurements for $EPS_1$. In both subfigures, the first row presents fidelity metrics, while the second row reports the corresponding entanglement-throughput metrics.

Specifically, from the reconstructed state, we calculate the Bell-state fidelity, that we call $F_{\mathrm{raw}}$, and the fidelity optimized over local unitary transformations, $F_{\mathrm{LU}}$. Their difference $\Delta F_{\mathrm{LU}} = F_{\mathrm{LU}}-F_{\mathrm{raw}}$ quantifies the locally correctable fidelity improvement and is expressed
in percentage points (pp).
The first row of each subfigure in Figure\ref{fig:fidelity} shows these three quantities. 
The small values of $\Delta F_{\mathrm{LU}}$ indicate that local rotations provide only a limited improvement and that most of the observed fidelity reduction cannot be removed through local basis corrections alone.
The second row of each subfigure of Figure~\ref{fig:fidelity} relates the measured source performance to the achievable entanglement throughput, defined as the amount of usable entanglement generated per unit time. Specifically, the first
panel reports the background-corrected detected pair rate
$R_{\mathrm{pair}}$, i.e., the number of photon pairs detected per second.
Reducing the VOA     significantly increases this rate\footnote{Again, as the multiphoton analysis, the reported values remain detection-level quantities and therefore include optical losses and detector inefficiencies.}.
The central panel reports the one-way coherent-information hashing lower bound, $D_{\rightarrow}^{\mathrm{hash,LB}}$, expressed in ebit per
detected pair. This quantity estimates the asymptotic distilled entanglement achievable using the one-way LOCC hashing protocol \cite{DevWin-05}. In contrast to $R_{\mathrm{pair}}$, it decreases at the operating points with the lowest VOA, revealing a trade-off between the number of detected pairs and their distillable entanglement.

The last panel combines these two contributions:
\begin{equation}
R_{\mathrm{ebit}}^{\mathrm{LB}}
=
R_{\mathrm{pair}}
D_{\rightarrow}^{\mathrm{hash,LB}}.
\end{equation}
This quantity represents a lower bound on the achievable entanglement throughput, i.e., the asymptotic number of useful ebits that could ideally be distilled per second from the detected photon pairs using one-way LOCC.
For $EPS_2$ (Figure~\ref{fig:fidelity_eps2}), the achievable entanglement throughput exhibits a clear trade-off between the detected pair rate and the distillable entanglement per pair. For example, a VOA of 18 dB provides approximately 0.91 ebit per pair, but its low detected pair rate limits the useful throughput to 0.5–0.9 kebit/s. Conversely, the maximum measured pair rate occurs at VOA $6$~dB and $50$~MHz, but the low hashing bound limits the corresponding throughput to $19.8$~kebit/s. The
largest value, approximately $61.4$~kebit/s, is obtained at VOA $6$~dB and $150$~MHz, which provides the most favorable balance between detected pair rate and distillable entanglement per pair.
The comparison between $EPS_1$ and $EPS_2$  reveals a clear trade-off between the quality of the detected pairs and their generation rate. Over the common operating region, $EPS_1$ generally exhibits higher Bell-state
fidelity and a larger one-way hashing lower bound, indicating a greater amount of distillable entanglement per detected pair. Conversely, $EPS_2$ 
provides a substantially higher detected pair rate at every common
operating point.

The higher rate of $EPS_2$  generally compensates for its lower
entanglement content per pair, leading to a maximum
$R_{\mathrm{ebit}}^{\mathrm{LB}}$ of approximately
$61.4$~kebit/s at $6$~dB and $150$~MHz. $EPS_1$ reaches a maximum of
approximately $46.6$~kebit/s at $3$~dB and $100$~MHz. Nevertheless, at
$6$~dB and $50$~MHz, $EPS_1$  provides a higher distillable-entanglement
rate despite its lower detected pair rate, owing to its substantially
larger hashing bound. These results demonstrate that neither fidelity nor
detected pair rate alone provides a complete source-level performance
metric; their combined contribution is captured by
$R_{\mathrm{ebit}}^{\mathrm{LB}}$.

\subsection{Temporal Stability of Entanglement Quality}

The stability of the entanglement source is evaluated by repeating the QST for fixed values of frequency and VOA. The analysis was performed for $EPS_2$, ITU channels 24–44, at 100 MHz and VOA 9 dB. Five consecutive measurements were collected over approximately 0.8 h.

Figure~\ref{fig:fidelity-temporal-stability} summarizes the temporal evolution of the Bell-state fidelity, the locally corrected fidelity, the one-way hashing lower bound, the detected pair rate, and the corresponding entanglement-throughput lower bound.

All measured quantities remain remarkably stable throughout the observation interval. The Bell-state fidelity remains centered at 95.665\% with a standard deviation of only 0.031\%, while the one-way hashing lower bound fluctuates around 0.7037 ebit per detected pair with a standard deviation of 0.0017 ebit per pair. The detected pair rate is similarly stable, averaging 41.71 kpair/s, which results in an entanglement throughput of approximately 29.35 kebit/s.

The fitted linear trends are negligible and do not indicate any systematic degradation of the source. The observed fluctuations are instead consistent with normal statistical and experimental variability. Although the present analysis is limited to five acquisitions over approximately 0.8 h and therefore assesses short-term repeatability rather than long-term stability, it indicates that the source operates consistently under fixed experimental conditions.

\subsection{Translation into Quantum-Memory Requirements}

While the one-way hashing lower bound $D_{\rightarrow}^{\mathrm{hash,LB}}$, introduced in Section~\ref{sec:Experimental reuslts}, quantifies the asymptotic distillable entanglement, the protocol requires a sufficiently large block of detected photon pairs before distillation can be performed.
We call $T_{\mathrm{acc}}$ the minimum time required to collect the $n$ detected entangled pairs forming the input block for one hashing round.
Specifically this time is given by:
\begin{equation}
T_{\mathrm{acc}}^{\mathrm{LB}}(n)
\simeq
\frac{n}{R_{\mathrm{pair}}}.
\label{eq:07}
\end{equation}
Assuming ideal and instantaneous local processing, the corresponding memory-coherence requirement is lower bounded by this time\footnote{In addition to this temporal requirement, each node must provide sufficient
storage capacity to retain its local halves of the $n$ entangled pairs processed by the protocol. Therefore, increasing the block size imposes requirements on both memory coherence time and multimode storage capacity.}:

\begin{equation}
T_{\mathrm{coh}}^{\mathrm{LB}}(n)
\simeq
T_{\mathrm{acc}}^{\mathrm{LB}}(n).
\label{eq:08}
\end{equation}
In other words, for a given input block size $n$, the experimentally measured detected-pair rate $R_{\mathrm{pair}}$ can be translated into a minimum quantum-memory coherence time required to accumulate the block before distillation.

Once the input block has been accumulated, the one-way hashing bound determines the asymptotic amount of distillable entanglement that can be extracted from it. Specifically, for an input block of n detected photon pairs, the corresponding number of distilled ebits is given by:
\begin{equation}
m_{\mathrm{ebit}}^{\mathrm{LB}}(n)
\simeq
nD_{\rightarrow}^{\mathrm{hash,LB}}.
\end{equation}
Therefore from Eq.~\ref{eq:07} and Eq.~\ref{eq:08}, the minimum quantum-memory coherence time required to produce $m$ distilled ebits becomes:
\begin{equation}
T_{\mathrm{coh}}^{\mathrm{LB}}(m)
\simeq
\frac{m_{\mathrm{ebit}}^{\mathrm{LB}}(n)}
{R_{\mathrm{pair}}D_{\rightarrow}^{\mathrm{hash,LB}}}=\frac{m_{\mathrm{ebit}}^{\mathrm{LB}}(n)}
{R_{\mathrm{ebit}}^{\mathrm{LB}}}.
\end{equation}
The resulting memory-coherence requirements are summarized in Figure~\ref{fig:memory-requirement}. At VOA 18 dB, the high hashing lower bound,$(D_{\rightarrow}^{\mathrm{hash,LB}}\approx0.91$ ebit/pair), is offset by the low detected pair rate $R_\mathrm{pair}$, limiting the achievable entanglement throughput to approximately 0.54-0.92 kebit/s and requiring up to 169 s to accumulate a block of $10^5$ detected pairs.

Reducing the VOA shortens the accumulation time at the expense of a lower distillable yield per pair. The maximum entanglement throughput is obtained at 6 dB and 150 MHz ($R_{\mathrm{ebit}}^{\mathrm{LB}}\approx61.4$  kebit/s), requiring approximately 9.62 ms and 0.962 s to accumulate $10^3$ and $10^5$ detected pairs, respectively. The shortest accumulation time is instead achieved at 6 dB and 50 MHz because of its higher detected pair rate, although its lower hashing bound reduces the resulting entanglement throughput.

These results further emphasize that maximizing Bell-state fidelity or distillable entanglement per detected pair does not necessarily maximize network-level performance. The optimal operating point depends on whether the primary objective is to maximize the achievable entanglement throughput or to minimize the quantum-memory coherence time required for block accumulation.

\begin{figure}[t]
    \centering
    \includegraphics[width=\columnwidth]
    {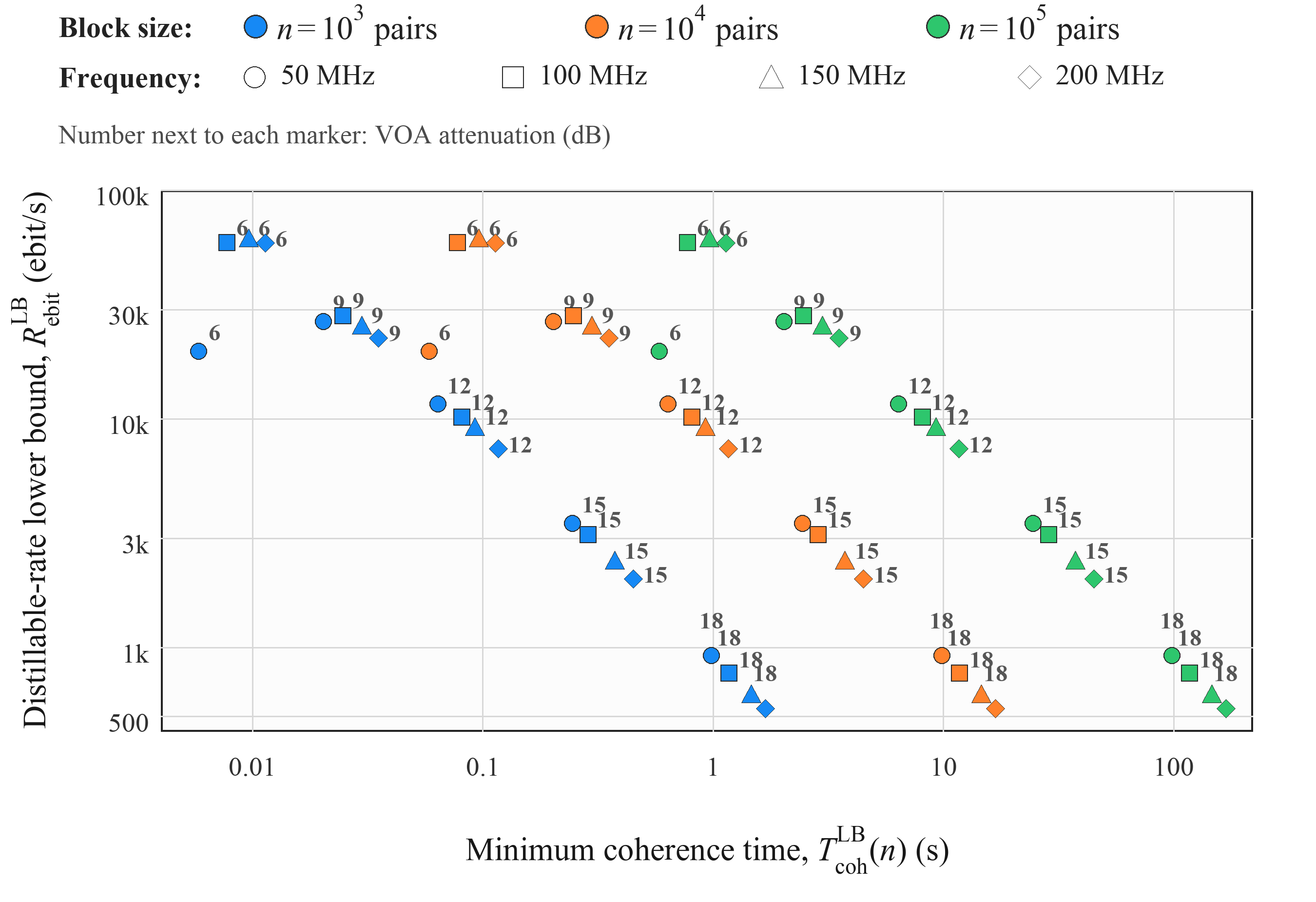}
    \caption{Memory-coherence lower bound versus one-way
    distillable-entanglement rate for $EPS_2$, channels 24--44.
    Colours identify the block size, marker shapes the pump frequency,
    and numerical labels the VOA in dB.}
    \label{fig:memory-requirement}
    \hrulefill
\end{figure}

\subsection{Discussion and Future Directions}

The reported coherence times represent ideal lower bounds rather than complete hardware requirements, as they rely on several assumptions regarding the implementation of entanglement distillation in future quantum-network nodes.

First, the analysis assumes the availability of quantum memories capable of preserving the stored photonic states for at least the estimated coherence times. Although significant progress has been achieved in quantum-memory technologies, current experimental platforms generally do not yet provide the combination of storage time required for large-scale photonic quantum networking. Consequently, the reported coherence times should be interpreted as target memory specifications rather than currently achievable performance \cite{HesKhaEng-16}.

Second, the achievable entanglement throughput is evaluated using the one-way hashing lower bound, which assumes ideal implementation of LOCC. In practice, realizing such distillation protocols for photonic qubits remains challenging. Most experimental demonstrations of entanglement distillation and quantum error correction have been developed in superconducting-qubit platforms, where high-fidelity local quantum gates are readily available. Extending these capabilities to photonic quantum networks requires efficient photonic quantum processing or reliable interfaces between optical photons, which are naturally suited for long-distance communication, and stationary quantum processors, many of which operate at microwave frequencies. The development of high-efficiency microwave-to-optical quantum transducers therefore remains an important technological challenge \cite{CalDavHan-25}.

Despite these assumptions, the proposed characterization extends the conventional description of entanglement sources beyond Bell-state fidelity and detected pair rate alone. By translating experimentally measured source performance into network-oriented quantities—including distillable-entanglement throughput, minimum memory-coherence time, and storage-capacity requirements—it provides metrics that are directly relevant to the engineering of future entanglement-based quantum networks.

Future work should incorporate experimentally characterized quantum memories, including finite storage efficiency and decoherence, realistic implementations of LOCC operations, finite-block-length distillation protocols, gate and measurement errors, classical-control latency, and the performance of practical photonic–matter interfaces. Such extensions would enable a more comprehensive assessment of the end-to-end performance achievable in realistic quantum-network architectures.

\bibliographystyle{IEEEtran}
\bibliography{bibliography.bib}

@inproceedings{ClaYanWanNejSimJos-24,
  title={What is the best wavelength for fibre quantum communication?},
  author={Clark, Marcus J and Yang, Ruizhi and Wang, Rui and Nejabati, Reza and Simeonidou, Dimitra and Joshi, Siddarth K},
  booktitle={Quantum 2.0},
  pages={QTh3A--43},
  year={2024},
  organization={Optica Publishing Group}
}

@article{YinRenLu-12,
  title={Quantum teleportation and entanglement distribution over 100-kilometre free-space channels},
  author={Yin, Juan and Ren, Ji-Gang and Lu, He and Cao, Yuan and Yong, Hai-Lin and Wu, Yu-Ping and Liu, Chang and Liao, Sheng-Kai and Zhou, Fei and Jiang, Yan and others},
  journal={Nature},
  volume={488},
  number={7410},
  pages={185--188},
  year={2012},
  publisher={Nature Publishing Group UK London}
}

@article{JakPreKau-14,
  title={Time-bin entangled photons from a quantum dot},
  author={Jayakumar, Harishankar and Predojevi{\'c}, Ana and Kauten, Thomas and Huber, Tobias and Solomon, Glenn S and Weihs, Gregor},
  journal={Nature communications},
  volume={5},
  number={1},
  pages={4251},
  year={2014},
  publisher={Nature Publishing Group UK London}
}

@article{AltJefKwi-05,
  title={Photonic state tomography},
  author={Altepeter, Joseph B and Jeffrey, Evan R and Kwiat, Paul G},
  journal={Advances in atomic, molecular, and optical physics},
  volume={52},
  pages={105--159},
  year={2005},
  publisher={Elsevier}
}

@article{RFC9340,
  title={RFC 9340: Architectural principles for a quantum internet},
  author={Kozlowski, Wojciech and Wehner, Stephanie and Van Meter, Rodney and Rijsman, Bruno and Cacciapuoti, Angela Sara and Caleffi, Marcello and Nagayama, Shota},
  year={2023},
  publisher={RFC Editor}
}

@article{CalDavHan-25,
  title={Quantum transduction: Enabling quantum networking},
  author={Caleffi, Marcello and D’Avossa, Laura and Han, Xu and Cacciapuoti, Angela Sara},
  journal={IEEE Communications Surveys \& Tutorials},
  year={2025},
  publisher={IEEE}
}

@article{CacCalTaf-19,
  title={Quantum internet: Networking challenges in distributed quantum computing},
  author={Cacciapuoti, Angela Sara and Caleffi, Marcello and Tafuri, Francesco and Cataliotti, Francesco Saverio and Gherardini, Stefano and Bianchi, Giuseppe},
  journal={IEEE Network},
  volume={34},
  number={1},
  pages={137--143},
  year={2019},
  publisher={IEEE}
}

@INPROCEEDINGS{TiaWuLi-24,
  author={Hu, Tianjie and Wu, Jindi and Li, Qun},
  booktitle={2024 IEEE 44th International Conference on Distributed Computing Systems (ICDCS)}, 
  title={Quantum Network Routing Based on Surface Code Error Correction}, 
  year={2024},
  volume={},
  number={},
  pages={1236-1247},
  doi={10.1109/ICDCS60910.2024.00117}}

@article{LiWanMin-22,
  title={Fidelity-guaranteed entanglement routing in quantum networks},
  author={Li, Jian and Wang, Mingjun and Xue, Kaiping and Li, Ruidong and Yu, Nenghai and Sun, Qibin and Lu, Jun},
  journal={IEEE Transactions on Communications},
  volume={70},
  number={10},
  pages={6748--6763},
  year={2022},
  publisher={IEEE}
}

@inproceedings{LiuLiWan-24,
  title={Linkselfie: Link selection and fidelity estimation in quantum networks},
  author={Liu, Maoli and Li, Zhuohua and Wang, Xuchuang and Lui, John CS},
  booktitle={IEEE INFOCOM 2024-IEEE Conference on Computer Communications},
  pages={1421--1430},
  year={2024},
  organization={IEEE}
}

@book{Loudon2000,
  author    = {Rodney Loudon},
  title     = {The Quantum Theory of Light},
  edition   = {3},
  publisher = {Oxford University Press},
  year      = {2000}
}

@book{WallsMilburn2008,
  author    = {D. F. Walls and Gerard J. Milburn},
  title     = {Quantum Optics},
  edition   = {2},
  publisher = {Springer},
  year      = {2008}
}

@article{KwiMatei-95,
  title = {New High-Intensity Source of Polarization-Entangled Photon Pairs},
  author = {Kwiat, Paul G. and Mattle, Klaus and Weinfurter, Harald and Zeilinger, Anton and Sergienko, Alexander V. and Shih, Yanhua},
  journal = {Phys. Rev. Lett.},
  volume = {75},
  issue = {24},
  pages = {4337--4341},
  numpages = {0},
  year = {1995},
  month = {Dec},
  publisher = {American Physical Society},
  doi = {10.1103/PhysRevLett.75.4337}
}

@incollection{Lvovsky2015,
  author    = {Alexander I. Lvovsky},
  title     = {Squeezed Light},
  booktitle = {Photonics, Volume 1: Fundamentals of Photonics and Physics},
  editor    = {David L. Andrews},
  pages      = {121--163},
  publisher = {John Wiley \& Sons, Ltd},
  year      = {2015},
  doi       = {10.1002/9781119009719.ch5}
}

@article{MeySilMig-20,
  title={Single-photon sources: Approaching the ideal through multiplexing},
  author={Meyer-Scott, Evan and Silberhorn, Christine and Migdall, Alan},
  journal={Review of Scientific Instruments},
  volume={91},
  number={4},
  year={2020},
  publisher={AIP Publishing}
}

@article{TalHesDav-26,
  title={Quantum entanglement distribution coexisting with high-rate, broadband classical optical communications over a real-world fiber connecting remote, synchronized nodes},
  author={Talcott, Gina M and Hess, Ahnnika I and d’Avossa, Laura and Kohlert, Scott J and Yeh, Fei I and Chen, Jim Hao and Mambretti, Joe J and Rambo, Tim M and Kanter, Gregory S and Thomas, Jordan M and others},
  journal={Optica Quantum},
  volume={4},
  number={4},
  pages={342--352},
  year={2026},
  publisher={Optica Publishing Group}
}

@book{nielsen2010quantum,
  title={Quantum computation and quantum information},
  author={Nielsen, Michael A and Chuang, Isaac L},
  year={2010},
  publisher={Cambridge university press}
}

@article{Bennett-96,
  title={Purification of noisy entanglement and faithful teleportation via noisy channels},
  author={Bennett, Charles H and Brassard, Gilles and Popescu, Sandu and Schumacher, Benjamin and Smolin, John A and Wootters, William K},
  journal={Physical review letters},
  volume={76},
  number={5},
  pages={722},
  year={1996},
  publisher={APS}
}

@article{HorHorHor-98,
  title={Mixed-state entanglement and distillation: Is there a “bound” entanglement in nature?},
  author={Horodecki, Micha{\l} and Horodecki, Pawe{\l} and Horodecki, Ryszard},
  journal={Physical Review Letters},
  volume={80},
  number={24},
  pages={5239},
  year={1998},
  publisher={APS}
}

@article{NatTanHad-12,
  title={Superconducting nanowire single-photon detectors: physics and applications},
  author={Natarajan, Chandra M and Tanner, Michael G and Hadfield, Robert H},
  journal={Superconductor science and technology},
  volume={25},
  number={6},
  pages={063001},
  year={2012},
  publisher={IOP publishing}
}

@article{DevWin-05,
  title={Distillation of secret key and entanglement from quantum states},
  author={Devetak, Igor and Winter, Andreas},
  journal={Proceedings of the Royal Society A: Mathematical, Physical and engineering sciences},
  volume={461},
  number={2053},
  pages={207--235},
  year={2005},
  publisher={The Royal Society}
}

@article{HesKhaEng-16,
  title={Quantum memories: emerging applications and recent advances},
  author={Heshami, Khabat and England, Duncan G and Humphreys, Peter C and Bustard, Philip J and Acosta, Victor M and Nunn, Joshua and Sussman, Benjamin J},
  journal={Journal of modern optics},
  volume={63},
  number={20},
  pages={2005--2028},
  year={2016},
  publisher={Taylor \& Francis}
}

@article{ZhoWanZou-20,
  title={Proposal for heralded generation and detection of entangled microwave--optical-photon pairs},
  author={Zhong, Changchun and Wang, Zhixin and Zou, Changling and Zhang, Mengzhen and Han, Xu and Fu, Wei and Xu, Mingrui and Shankar, Shyam and Devoret, Michel H and Tang, Hong X and others},
  journal={Physical review letters},
  volume={124},
  number={1},
  pages={010511},
  year={2020},
  publisher={APS}
}

@article{LuLisGae-23,
  title={Frequency-bin photonic quantum information},
  author={Lu, Hsuan-Hao and Liscidini, Marco and Gaeta, Alexander L and Weiner, Andrew M and Lukens, Joseph M},
  journal={Optica},
  volume={10},
  number={12},
  pages={1655--1671},
  year={2023},
  publisher={Optica Publishing Group}
}

\end{document}